\documentstyle[twoside,fleqn,hyperon,amsfonts,amssymb,psfig,here]{article}
\begin{document}
\title{
\vspace*{-1cm}
The Search for Pentaquark Baryon with Hidden Strangeness \\
{\large 
\it Talk on HYPERON99, Fermilab, September, 27-29, 1999}
}

\author{L.G.Landsberg\\
Institute for High Energy Physics,
Protvino, Moscow region, 142284, Russia}

\begin{abstract}
Evidences for new baryon states with mass $>$1.8~GeV were obtained in the
experiments of the SPHINX Collaboration in studying hyperon-kaon mass
spectra in several proton diffractive reactions. The main result of these
experiments is the observation of $X(2000) \rightarrow \Sigma K$ state with
unusual dynamical features (narrow width, anomalously large branching
ratios for the decay channels with strange particle emission). The
possibility of the interpretation of this state as cryptoexotic
pentaquark baryon with hidden strangeness is discussed. The additional
data which are supported the real existence of $X(2000)$ baryon are
also presented.
\end{abstract}
\maketitle

Extensive studies of the diffractive baryon production and search for
cryptoexotic pentaquark baryons with hidden strangeness ($B_\phi = |qqqs \bar s>$;
here $q=u,~d$ quarks) are being carried out by the SPHINX Collaboration at
IHEP accelerator with 70~GeV proton beam. This program was described in detail
in reviews~[1].

The cryptoexotic $B_\phi$ baryons do not have external exotic quantum numbers
and their complicated internal valence quark  structure can be established only
indirectly, by examination of their unusual dynamic properties which
are quite different from those for ordinary $|qqq >$ baryons. Examples of such
anomalous features are as listed below (see~[1] for more details):

1. The dominant OZI allowed decay modes of $B_\phi$ baryons are the ones with
strange particles in the final state (for ordinary baryons such decays have branching ratios
at the per cent level).

2. Cryptoexotic $B_\phi$ baryons can possess both large masses ($M>1.8-2.0$~GeV)
and narrow decay widths ($\Gamma \leq 50-100$~MeV). This is due to a
complicated internal color structure of these baryons with significant quark
rearrangement of color clusters in the decay processes and due to a limited
phase space for the OZI allowed $B \rightarrow YK$ decays. At the same time,
typical decay widths for the well established $|qqq >$ isobars with similar
masses are $\geq 300$~MeV.

As was emphasized in a number of papers (see reviews~[1,2]), diffractive
production processes with Pomeron exchange offer new tools in searches for the
exotic hadrons. Originally, the interest was concentrated
on the model of Pomeron with small cryptoexotic $(qq \bar q \bar q)$ component.
In modern notions Pomeron is a multigluon system which allows for
production of the exotic hadrons in gluon-rich diffractive processes.

The Pomeron exchange mechanism in diffractive production reactions can induce
the coherent processes on the target nucleus. In such processes the nucleus
acts as a whole. Owing to the difference in the absorptions of single-particle
and multiparticle objects in nuclei, coherent processes could serve as an
effective tool for separation of resonance against non-resonant multiparticle
background.

In previous measurement on the SPHINX setup several unusual baryonic states
were observed in the study of coherent
diffractive production reactions
\begin{equation}
p+N(C) \rightarrow [\Sigma^0 K^+] + N(C)
\end{equation}
and
\begin{equation}
p+N(C) \rightarrow [\Sigma^*(1385)^0K^+]+N(C)
\end{equation}
(see [1,2] and the references therein; here $C$ corresponds to coherent
reaction on carbon nuclei):

a) the state $X(2000)^+ \rightarrow \Sigma^0 K^+$ with mass $M=1997\pm7$~MeV
and the width $\Gamma = 91 \pm 17$~MeV;

b) the state $X(1810)^+ \rightarrow \Sigma^0 K^+$ with $M=1812 \pm 7$~MeV and $\Gamma=56\pm16$~MeV;

c) the state $X(2050)^+ \rightarrow \Sigma^*(1385)^0K^+$ with $M=2052\pm 6$~MeV
and $\Gamma=35^{+22}_{-35}$~MeV
(preliminary data obtained in the old run; new data are now under analysis).

The states $X(1810)$ and $X(2050)$ are seen only in the region of very small
$P^2_T(\lesssim 0.01-0.02~{\rm GeV}^2)$. The states $X(2000)$ and $X(2050)$
have anomalously large branching ratios for decay channels with strange particle
emission
\noindent
$$
R=BR[X(2000);~X(2050) \rightarrow YK]/BR(X(2000),
$$
\begin{equation}
X(2050) \rightarrow p \pi^+ \pi^-;~\Delta^{++} \pi^-)
\gtrsim 1 \div 10.
\end{equation}

This feature and their comparatively narrow decay widths make these states
good candidates for exotic baryons with hidden strangeness.

In what follows we present the results of a new analysis~[3] of the data
obtained in the run with partially upgraded SPHINX spectrometer where
conditions for $\Lambda$ and $\Sigma^0$ separation were greatly
improved as compared to old version of this setup (see~[4]). The key element
of a new analysis consists in detailed study of the $\Sigma^0 \rightarrow
\Lambda + \gamma$ decay separation. New analysis gave possibility to increase statistics more
than in two times. Detailed GEANT Monte-Carlo simulation was used for efficiency
calculations and cross section estimations.

The effective mass spectrum $M(\Sigma^0 K^+)$ in (1) for all $P^2_T$ is
presented in Fig.1. The peak of $X(2000)$ baryon state with
$M=1986\pm 6$~MeV and $\Gamma=98\pm20$~MeV is seen very clearly in this
spectrum with a good statistical significance. Thus, the reaction
\vspace*{-0.1cm}
\begin{eqnarray}
p+N \rightarrow &\!&  X(2000) + N,
\label{eq:4a}\\
&\!&~\protect \raisebox {1.11ex}{$ \lfloor$} \! \! \!
\rightarrow \Sigma^0 K^+ \nonumber
\end{eqnarray}
is well separated in the SPHINX data. We estimated the cross section for
$X(2000)$ production in (4):
\begin{equation}
\begin{array}{l}
\sigma[p + N \rightarrow X(2000) + N] \cdot BR[X(2000) \rightarrow \nonumber \\
\rightarrow \Sigma^0 K^+] = 95 \pm 20~{\rm nb/nucleon}
\end{array}
\end{equation}
(with respect to one nucleon under the assumption of $\sigma \propto A^{2/3}$,
e.g. for the effective number of nucleons in carbon nucleus equal to 5.24).
The parameters of $X(2000)$ peak are not sensitive to different photon cuts.

The $dN/dP^2_T$ distribution for reaction (4) is shown in Fig.2. From this distribution
the coherent diffractive production reaction on carbon nuclei is
identified as a diffraction peak with the slope $b \simeq 63\pm 10$~GeV$^{-2}$.
The cross section for coherent reaction is determined as

$$\sigma[p+C \rightarrow X(2000)^+ + C]_{{\rm Coherent}} \cdot$$
$$\cdot BR[X(2000)^+ \rightarrow \Sigma^0 K^+] = $$
\setcounter{equation}{5}
\begin{equation}
\quad \quad \quad \quad =260 \pm~60~{\rm nb/C~nuclei}.
\end{equation}

The errors in the values of (5) and (6) are statistical only.
Additional systematic errors are about $\pm 20$\% due to uncertainties in the
cuts, in the Monte Carlo efficiency calculations and in the absolute
normalization.

In the mass spectrum $M(\Sigma^0 K^+)$ in Fig.1 there is only
a slight indication for X(1810) structure which was observed earlier in the
study of coherent reaction (1). This difference is caused by a large
background in this region for the events in Fig.1 (for all $P^2_T$
values).

\begin{figure}[H]
\psfig{file=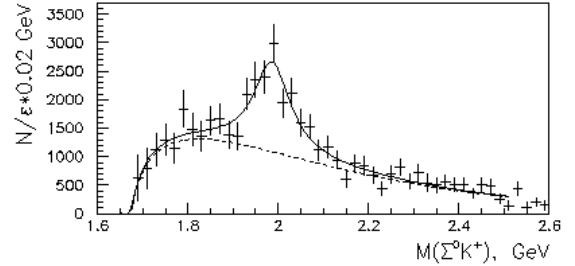,width=7.5cm}
\caption{Invariant mass spectrum $M(\Sigma^0K^+)$ in diffractive reaction
$p+N\to[\Sigma^0K^+]+N$ for all $P^2_T$ (weighted with the efficiency of the 
setup). The peak $X(2000)$ with parameters $M=1986\pm 6\,{\rm MeV}$ and
$\Gamma = 98\pm 20\,{\rm MeV}$ is clearly observed in this spectrum with a 
very high statistical significance.}
\end{figure}

\begin{figure}[H]
\psfig{file=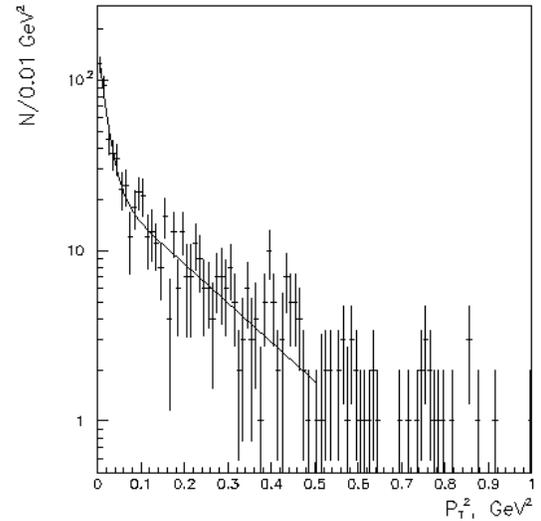,width=7.5cm}
\caption{$dN/dP^2_T$ distribution for the diffractive production
reaction $p+N \to X(2000)+N$. The distribution is fitted in the form 
$dN/dP^2_T = a_1 exp(-b_1 P^2_T)+a_2 exp(-b_2 P^2_T)$ with parameters
$b_1 = 63 \pm 10 \,{\rm GeV}^{-2}$,
$b_2 = 5.8\pm 0.6\,{\rm GeV}^{-2}$.
}
\end{figure}

But in the new data for coherent reaction (1) in the mass spectra $M(\Sigma^0 K^+)$
both states $X(2000)$ and $X(1810)$ are clearly seen. Study of the yield of $X(1810)$ as
function of $P^2_T$ demonstrates that this state is produced only in
the region of very small $P^2_T$ ($\lesssim 0.01$~GeV$^2$) where it is well
defined (see Fig.3). From this data parameters of $X(1810)$ are determined
\arraycolsep 1pt
{\begin{equation}
X(1810) \rightarrow \Sigma^0 K^+\ \left\{ \begin{array}{ll}
M & =  1807 \pm \enspace7~{\rm MeV} \\[3mm]
\Gamma&  =  \enspace \enspace 62 \pm 19~{\rm MeV},
\end{array} \right.
\end{equation}}
as well as the coherent cross section
\noindent
$$
\sigma[p+C \rightarrow X(1810)+C]_{p^2_T<0.01~{\rm GeV}^2} \cdot$$
$$\cdot  BR[X(1810) \rightarrow
\Sigma^0 K^+]=$$
$$
=215\pm 44~{\rm nb}~(\pm 30\%~{\rm syst.}). \eqno(8) $$

\begin{figure}[H]
\psfig{file=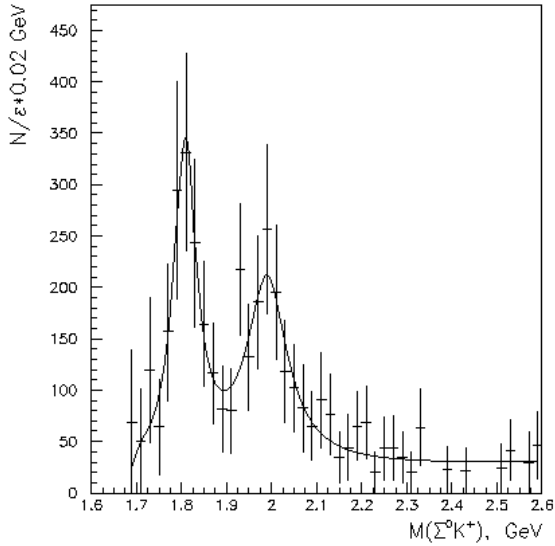,width=7.5cm}
\caption{Invariant mass spectrum $M(\Sigma^0K^+)$ in the coherent diffractive 
production reaction
$p+C\to[\Sigma^0K^+]+C$ in the region of very small $P^2_T<0.01\,{\rm GeV}^2$ 
(weighted with the setup efficiency).
In this region $X(1810)$ peak with parameters $M=1807\pm 7\,{\rm MeV}$ and
$\Gamma = 62\pm 19\,{\rm MeV}$ is clearly seen.
}
\end{figure}

To explain the unusual properties of $X(1810)$ state in a very small
$P^2_T$ region, the hypothesis of the electromagnetic production
of this state in the Coulomb field of carbon nucleus was proposed~[5]
and it seems to be in no contradictions with the experimental data
for the coherent cross section (8)~--- see [3]. This
hypothesis  is also supported by observation of $\Delta(1232)^+$ Coulomb
production on carbon nuclei in the SPHINX experiment~[5].

The data on $X(2000)$ baryon state with unusual dynamical
properties (large decay branching with strange particle emission,
limited decay width) were obtained with a good statistical
significance in the different SPHINX runs with widely different experimental
conditions and for several kinematical regions of reaction (1). The average values
of the mass and width of $X(2000)$ state (for different kinematical regions 
and cuts) are
\setcounter{equation}{8}
\begin{equation}
X(2000) \rightarrow \Sigma^0 K^+\ \left\{ \begin{array}{cc}
M & =  1989 \pm \enspace6~{\rm MeV} \\[3mm]
\Gamma & =  \enspace \enspace 91 \pm 20~{\rm MeV}
\end{array} \right.
\end{equation}

Due to its anomalous properties the $X(2000)$ state can be considered
as a serious candidate for pentaquark exotic baryon with
hidden strangeness: $|X(2000)> = |uud s \bar s>$.
Recently we have obtained some new additional data to support the
reality of $X(2000)$ state.

1. In the experiments with the SPHINX setup we studied the reaction
\begin{eqnarray}
p+N(C)\rightarrow &\!& [\Sigma^+ \hspace*{7mm}  K^0] + N(C). \\
&\!&\hspace{2mm}^|\hspace{-2mm} \rightarrow p \pi^0
\hspace{2mm}^|\hspace{-2mm} \rightarrow \pi^+ \pi^-\nonumber
\end{eqnarray}
In spite of a limited statistics, we observed the $X(2000)$ peak and the
indication for $X(1810)$ structure in this reaction which
are quite compatible with the data for reaction (1)~[6].

2. In the experiment at the SELEX (E781) spectrometer~[7]
with the $\Sigma^-$ hyperon beam of the Fermilab Tevatron, the
diffractive production reaction
\begin{equation}
\Sigma^- + N \rightarrow [\Sigma^- K^+ K^-]+N
\end{equation}
was studied~at~the~beam mo\-mentum $P_{\Sigma^-} \simeq$600~GeV.
In the invariant mass spectrum $M(\Sigma^- K^+)$ for this
reaction a peak with parameters $M=1962 \pm 12$~MeV and $\Gamma= 96 \pm 32$~MeV
was observed (see Fig.4 and [8]). The parameters of this structure
are very close to the parameters of $X(2000) \rightarrow \Sigma^0 K^+$
state which was observed in the experiments at the SPHINX spectrometer. Thus, the
real existence of $X(2000)$ baryon seems to be supported by the data
from another experiment and in another process.

\begin{figure}[H]
\psfig{file=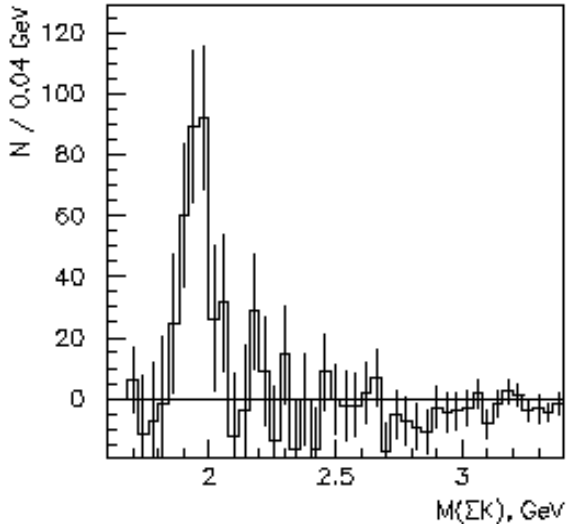,width=7.5cm}
\caption{Invariant mass spectrum $M(\Sigma^-K^+)$ in diffractive production
reaction
$\Sigma^-+N\to[\Sigma^-K^+] K^-+N$ (after background subtraction --
see [8]). In this spectrum the peak with parameters $M=1962\pm 12\,{\rm MeV}$ 
and $\Gamma = 96\pm 32\,{\rm MeV}$ (which are very near to the
parameters of $X(2000)$ peak in Fig.1) is observed.
}
\end{figure}

\section*{Conclusion}
In the study of diffractive production proton reactions with the SPHINX
setup we observed several interesting objects with anomalous properties.
The most important data were obtained for a new baryon state
$X(2000) \rightarrow \Sigma K$. Unusual features of this massive state
(relatively narrow decay width, large branching ratio for decay channels
with strange particle emission) make it a serious candidate for
cryptoexotic pentaquark baryon with hidden strangeness. We hope to
increase significantly our statistics in the near future
and to obtain a new information about the supposed exotic baryons.

\end{document}